\documentclass[conference]{IEEEtran}
\usepackage[pdftex]{graphicx}
\usepackage{graphicx}
\usepackage{booktabs}
\usepackage{xparse}
\usepackage{url}
\usepackage{subcaption}
\usepackage{multicol}
\NewDocumentCommand{\rot}{O{45} O{1em} m}{\makebox[#2][l]{\rotatebox{#1}{#3}}}%

\begin{document}

\title{SDN for End-Nodes: \\Scenario Analysis and Architectural Guidelines}

\author{\IEEEauthorblockN{Alberto Rodriguez-Natal\IEEEauthorrefmark{1}\IEEEauthorrefmark{2},
Vina Ermagan\IEEEauthorrefmark{1}, 
Kien Nguyen\IEEEauthorrefmark{3},
Sharon Barkai\IEEEauthorrefmark{4},\\
Yusheng Ji\IEEEauthorrefmark{5},
Fabio Maino\IEEEauthorrefmark{1} and
Albert Cabellos-Aparicio\IEEEauthorrefmark{2}}\\
\IEEEauthorblockA{\IEEEauthorrefmark{1}Cisco, USA. \IEEEauthorrefmark{3}NICT, Japan. \IEEEauthorrefmark{4}Fermi Serverless, USA. \IEEEauthorrefmark{5}NII, Japan. \IEEEauthorrefmark{2}BarcelonaTech, Spain.}}

\maketitle

\begin{abstract}
The advent of SDN has brought a plethora of new architectures and controller designs for many use-cases and scenarios. Existing SDN deployments focus on campus, datacenter and WAN networks. However, little research efforts have been devoted to the scenario of effectively controlling a full deployment of end-nodes (e.g. smartphones) that are transient and scattered across the Internet. In this paper, we present a rigorous analysis of the challenges associated with an SDN architecture for end-nodes, show that such challenges are not found in existing SDN scenarios, and provide practical design guidelines to address them. Then, and following these guidelines we present a reference architecture based on a decentralized, distributed and symmetric controller with a connectionless pull-oriented southbound and an intent-driven northbound. Finally, we measure a proof-of-concept deployment to assess the validity of the analysis as well as the architecture.
\end{abstract}

\IEEEpeerreviewmaketitle

\section{Introduction}
\label{intro}


In the recent years, Software Defined Networking (SDN) \cite{kreutz2015software} has been applied to different scenarios, namely campus, datacenter and WAN networks \cite{koponen2010onix,kreutz2015software}. However, little research efforts have been devoted to scenarios where the \emph{controlees} are end-nodes, e.g. smartphones, personal computers, home routers, etc. In this paper, we use the term end-nodes loosely to generally refer to end-user devices, which have not been traditionally considered in SDN architectures. Despite the scarce literature on the topic, we believe that an SDN architecture designed for end-nodes enables interesting use-cases, since companies and operators can push policies and control the traffic from its very origin.

This new SDN scenario brings a set of challenges that are not found in existing SDN deployments or that are strongly exacerbated in the end-nodes case. First, the number of controlees is very large and they are typically scattered through various networks, this imposes very high scalability and availability requirements to the controller. Second, this scenario comes with a high churn, since end-nodes are transient and mobile, and might connect and disconnect randomly \cite{falaki2010diversity}. Finally, the architecture has to face an additional complexity burden to operate the data-plane. As opposed to transit routers and switches, end-nodes do not aggregate traffic and thus, require fine-grain policies.

In this paper we analyze the challenges of deploying SDN for end-nodes and discuss practical design guidelines. As a result, we present a reference architecture based on a decentralized and symmetric controller with an intent-driven northbound and a pull-based southbound. We discuss the implementation options and build a proof-of-concept to show the validity of the analysis and the feasibility of the architecture. Finally, we summarize the relevant related work. 

\section{Use-cases}
\label{usecases}

SDN-capable end-nodes entitle network operators to extend the fine-grain control and centralized management provided by SDN to the very edge of the network. On one hand, this allows instantiating new services, and on the other, it eases the deployment and operation of existing ones. Some examples of use-cases that are enabled or enhanced via SDN for end-nodes are summarized below:


\paragraph*{Source control} SDN capabilities at end-nodes enable dynamic and remote control of their interfaces, which allows centralized management for load balancing and bandwidth aggregation. An operator would be able to balance a smartphone's traffic between Wi-Fi and cellular interfaces based on business agreements, remaining battery and/or device's geo-location (e.g. avoid traffic over cellular network when abroad). A remote backup company can leverage on multi-homed end-nodes to aggregate the bandwidth of all available interfaces and speed up transmissions.

\paragraph*{Destination control} SDN for end-nodes enables simpler and fine-grain control on the traffic destination. For instance, a content provider can detour traffic to one datacenter or another based on central control per end-node basis (e.g. to offer personalized content per end-user). Similarly, another company can offer remote VPN management on the devices of its employees dynamically tunneling the traffic depending on the destination (e.g. to decide on-the-fly which traffic should be encrypted).

\paragraph*{Path control} Path control can be achieved by combining SDN-capable end-nodes with some well-placed SDN nodes on transit networks. This allows companies to steer traffic through transit SDN nodes to, for instance, optimize routing or apply in-path functions. As an example, a parental filtering company can make use of such capabilities to filter traffic going to the end-nodes and block inappropriate content.

\section{Challenges}

SDN support at the end-nodes is not straightforward since it presents a set of new challenges when compared to existing SDN deployments. First of all, there is a \emph{high number} of controlees, potentially reaching millions of nodes. Although existing SDN deployments may handle a large number of controlees, this burden is heightened in the end-node scenario when combined with the rest of its challenges.

In addition, in existing SDN deployments controlees are physically connected to the SDN network, however in the end-nodes scenario the controlees are attached to \emph{legacy networks} at the edge. Such networks are usually out of the administrative control of the SDN domain. Moreover, the controlees are \emph{scattered} over these edge networks, potentially at a worldwide scale. 

Furthermore, end-nodes are \emph{transient}, highly mobile and may have intermittent connectivity with a central controller. This makes a difference with regular SDN where typically controlees are stationary and always available. Besides, and contrary to common SDN devices, end-nodes usually have \emph{less resources} available and are subject to high \emph{heterogeneity} in terms of capabilities and conditions (e.g. geo-location, spectrum usage, remaining battery, etc). Finally, end-nodes have to face \emph{lower traffic locality} in contrast to typical SDN devices that usually aggregate traffic and take advantage of this to speedup data-plane performance using caches, wildcards and/or longest prefix matches.

\section{Design Guidelines}
\label{guidelines}

\begin{table}[!t]
\begin{center}
\begin{tabular}{ |c|c|c|c|c|c|c|l| } 

  \multicolumn{1}{c}{\rot[25]{In-place networks}}
  & \multicolumn{1}{c}{\rot[25]{High \# of controlees}}
  & \multicolumn{1}{c}{\rot[25]{Scattered nodes}}
  & \multicolumn{1}{c}{\rot[25]{Low traffic locality}}
  & \multicolumn{1}{c}{\rot[25]{Transient devices}}
  & \multicolumn{1}{c}{\rot[25]{Heterogeneity}}
  & \multicolumn{1}{c}{\rot[25]{Scarce resources}}
    \\ 
  \midrule
 $\bullet$& & & & &$\bullet$& & Overlay deployment  \\ 

  &$\bullet$& &$\bullet$& & & & Scale-out architecture \\ 

 & &$\bullet$&&$\bullet$& & & Decentralized/symmetric  \\ 

  &$\bullet$&$\bullet$&$\bullet$&$\bullet$& & $\bullet$& Connectionless pull-based\\ 
  
    &$\bullet$&&$\bullet$&$\bullet$& & & Control-State decoupling  \\ 

  &$\bullet$&&&$\bullet$&$\bullet$& & Intent-driven  \\ 
  
  &$\bullet$& &$\bullet$& & &$\bullet$& Less granularity  \\ 
  \bottomrule
\end{tabular}
\end{center}
\caption{\textbf{Challenges to design principles} }
\label{tab:principles}
\end{table}

The resulting design principles to overcome the scenario challenges are summarized in Table \ref{tab:principles}. First, an \emph{overlay approach} that encapsulates traffic at the data-plane is required to bypass in-place networks and to offer a homogeneous view of the heterogeneous controlees. To be able to encapsulate overlay traffic into underlay packets, end-nodes have to retrieve the appropriate state from the controller's NIB (Network Information Base), term that we borrow from \cite{koponen2010onix}. Due to the large number of data-plane nodes and the low traffic locality, this generates a high ratio of requests from the southbound to the NIB and therefore, the controller must be able to \emph{scale-out} to accommodate this very high volume of messages. Furthermore, since end-nodes are scattered and mobile and southbound queries may reach any controller node, the controller should remain \emph{decentralized and symmetric}, i.e. all controller nodes should fulfill the same role and be disjoint, independent and interchangeable. 

Generally in SDN, data-plane nodes establish a connection with the controller to exchange state. End-nodes are transient and mobile, thus a regular connection based southbound protocol would impose frequent connection reestablishment. As a consequence an important design guideline is to use a \emph{connectionless southbound} that mitigates this burden and takes advantage of the symmetry of the controller to decouple controlees from specific controller nodes. This eases end-nodes roaming and also simplifies on-demand controller nodes instantiation (i.e. a new controller node can transparently start serving end-nodes previously served by another). On the other hand, the controller cannot push the state to transient nodes, since they might be not available at the time the state is generated. Furthermore, they can be more constrained in resources and thus may need to minimize the state they store. To address these scenario-specific constraints, end-nodes should use a data-driven \emph{pull-based mechanism} for the southbound. That allows them retrieving the state on demand and only for the actual traffic they need to forward. The pull-based approach can be complemented with push notifications for state updates and default forwarding policies for the periods when the state cannot be retrieved. To facilitate the southbound requirements, the controller should enforce a strong \emph{decoupling of control and state}. Clear distinction of the control policies and the forwarding state is required in order to easily expose the latter to the data-plane nodes. 

In addition, as opposed to transit routers and switches, end-nodes do not aggregate traffic. Flow policies need to be installed at the end-nodes themselves, leading to more frequent queries to the controller. Also, reduced locality renders caching or prefetching schemes less effective since each node has to fetch its own policies. In this context it is not advisable to follow the common multi-tuple lookup granularity used in SDN deployments, and rather it is recommended to use \emph{less granular} policies as much as possible (e.g. IP granularity). This simplifies data-plane implementation and complexity at the NIB.


Furthermore, the very large number of nodes to control, as well as the fact that they are mobile and transient, makes programming via imperative northbound interfaces very complex. Moreover, the heterogeneity of the devices and their network conditions makes challenging to define per end-node policies. As a result, we suggest to use an \emph{intent-driven approach} (i.e. a declarative language \cite{foster2013languages,kim2013improving},) as the northbound interface, to define abstract policies to apply to data-plane traffic. 

\begin{figure}[!t]
\centering
\includegraphics[width=0.45\textwidth]{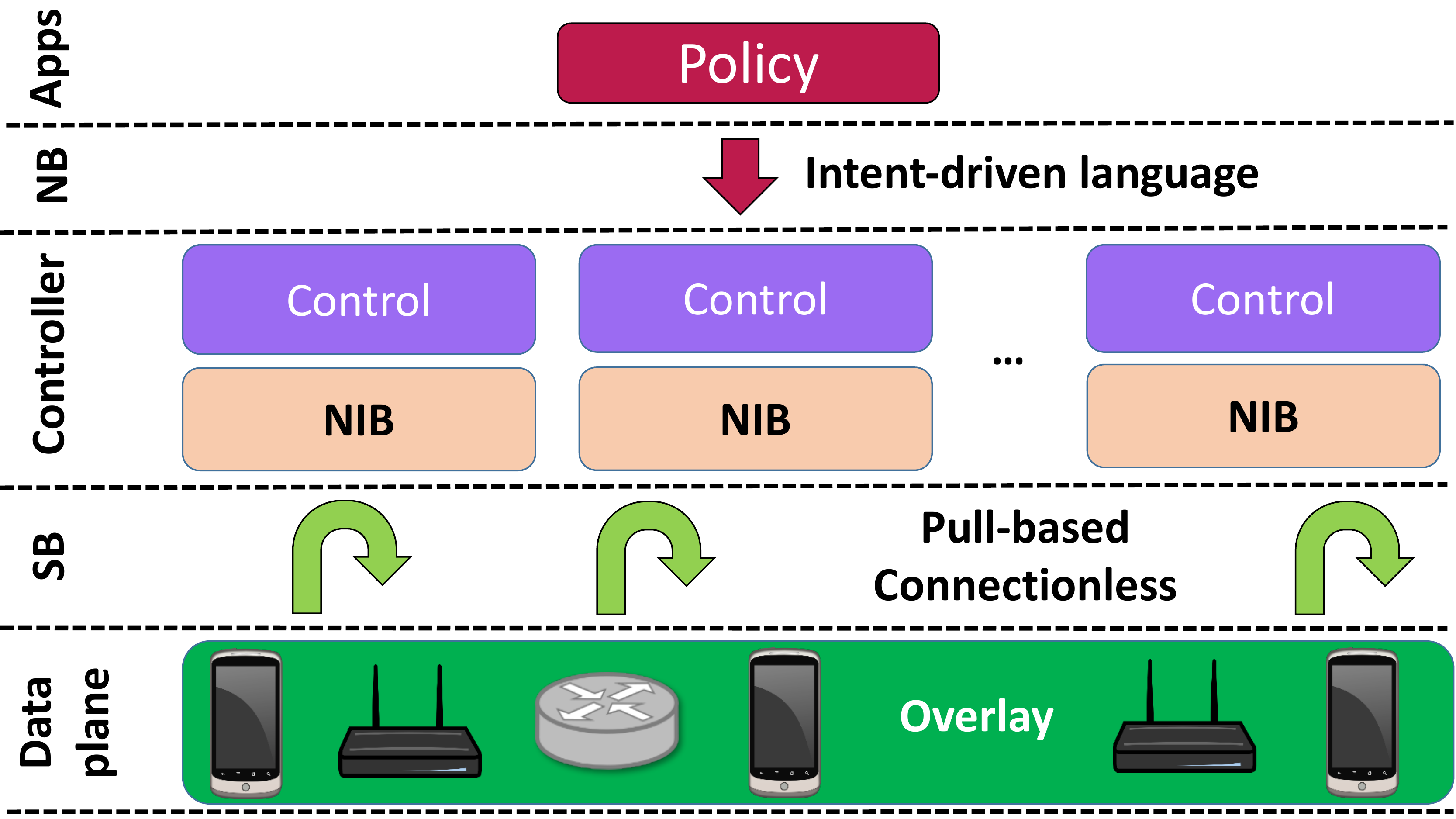}
\caption{\textbf{Architecture} }
\label{fig:arch}
\end{figure}

\section{Architectural components}
\label{arch}

Following the design principles from Section \ref{guidelines}, we propose the reference architecture summarized in Fig. \ref{fig:arch}. The controller is divided into two clear components, control and NIB. Abstract intent-driven policies are rendered into specific data-plane state by the control module and stored in the NIB. The NIB is then accessed on demand by the data-plane nodes using a pull-based connectionless southbound. The data-plane nodes are provisioned with encapsulation capabilities to join a common overlay. Implementation options for the architectural components discussed below can be found in section \ref{impl}.




\subsection{Northbound}
\label{northbound}

The policies are enforced from the northbound using a declarative intent-driven language. The northbound language is rendered at the controller to generate the data-plane state stored in the NIB. In particular, we propose to use a declarative language leveraging on groups, given that related nodes will likely share the same policies. 

Using a group-oriented language, end-nodes belong to groups and group policies define how to prioritize links, balance traffic or chain hops on the underlay. The policies can be defined for a pair of groups, when one sends traffic to the other, or in general for the group's default ingress and egress policies, thus avoiding defining all possible group pairs. For instance, a particular use-case can define the groups \emph{Employees' Laptops} and \emph{West Coast Office} with the following default policies, \emph{Employees' laptops} should use \emph{Ethernet} for egress traffic and \emph{West Coast Office} ingress traffic should be balanced between \emph{Link-1} and \emph{Link-2}. However, the use-case may also define that when the traffic from \emph{Employees' Laptops} goes to \emph{West Coast Office} it should use \emph{LTE} as egress, pass through hop \emph{Firewall} and use \emph{Link-3} as ingress. More complex groups are possible, such \emph{Employees' Laptops with battery below 15\%}, however complex use-cases require careful design.

\subsection{Controller}
\label{controller}

\begin{figure}[!t]
\centering
\includegraphics[width=0.5\textwidth]{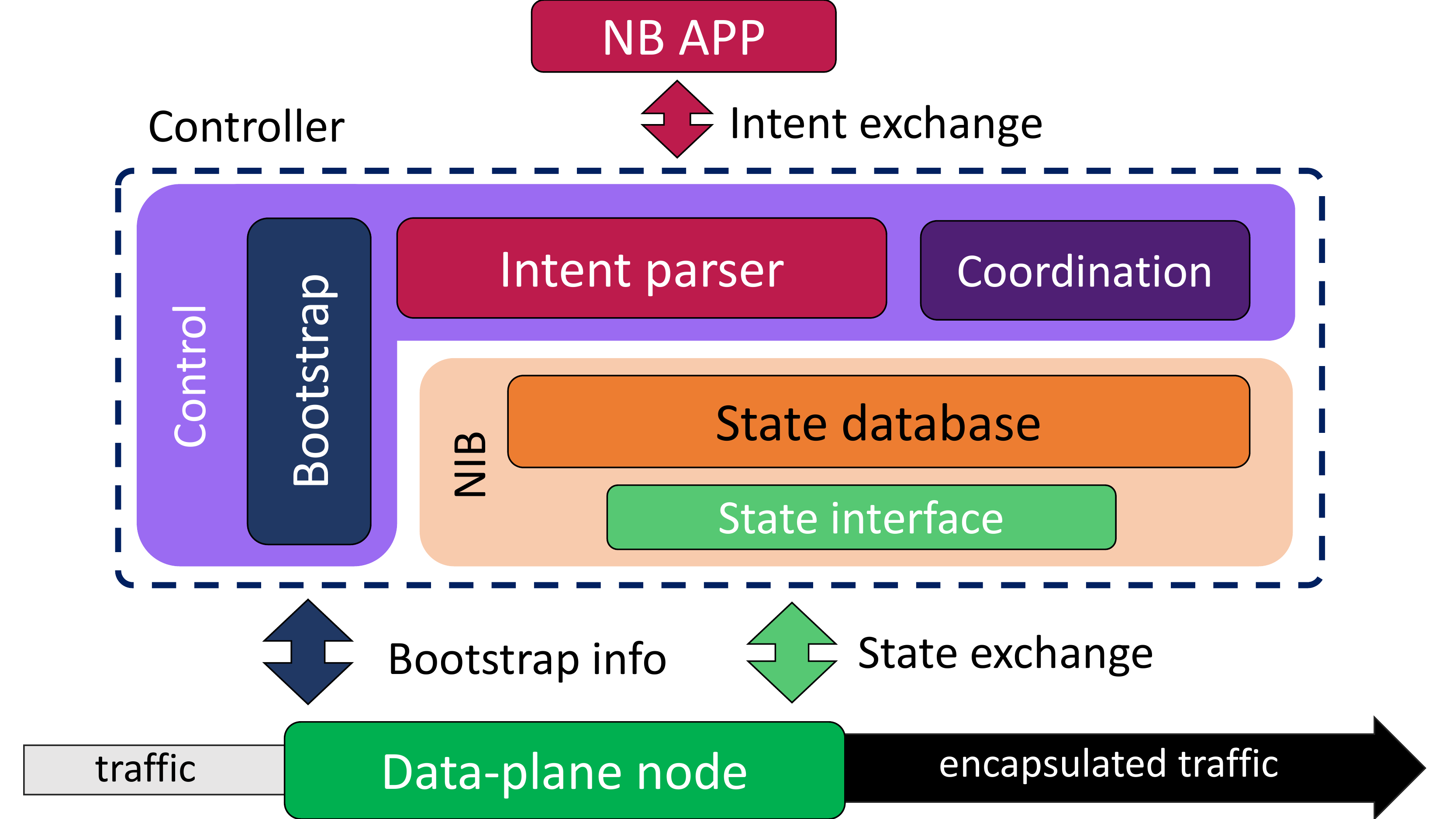}
\caption{\textbf{Controller node detail} }
\label{fig:node}
\end{figure}

The controller is supported by distributed, decentralized and symmetric nodes as the one depicted in Fig. \ref{fig:node}. They have an intent parser module that processes the northbound polices that result in NIB state. The bootstrap module is used to initialize data-plane nodes, while the coordination module allows the distributed controller modules to operate together. Via a southbound protocol each controller node exchanges state between the NIB and the southbound nodes. The control and NIB components are loosely coupled and can be scaled independently. For example, in a use case with light intent rendering but heavy southbound state retrieval, a single control module may handle several NIB partitions at a single controller location.


\paragraph*{Network Information Base (NIB)}

The NIB stores the rendered state generated by the intent parser and exposes it to the data-plane nodes.  For the scenario considered, the NIB does not need to represent a complete view of the network topology, just the relations and interactions among the groups and the nodes belonging to them.  In its very basic form, the state stored at the NIB should let the data-plane nodes know which source and destination addresses to use when encapsulating overlay traffic into underlay packets. The specific state is highly dependent on the particular use-case, with more complex use-cases requiring more complex state.

In that sense, the NIB should allow for a certain degree of generality to allow different use-cases and support highly heterogeneous end-devices. It should be noted that in the end-node scenario it is feasible to keep high generality without comprising system performance. This is possible since the scenario does not require a real-time detailed view of the network (state transactions impose less burden) and end-nodes do not need to react as fast as typical SDN equipment to state changes (more complex state processing is possible).

However, complex use-cases result in more database queries and potentially less parallelization of such queries. The consequence is increased latency per request and more NIB nodes needed. For instance, a use-case where groups depend on remaining battery and current geo-location makes necessary to obtain first the current group for an end-node prior to query for group-based path policies. The architecture has to keep a balance between the complexity of the queries (defined by the use case), the number of southbound requests (defined by the traffic and its locality), the maximum southbound latency accepted and the number and capacity of the NIB partitions deployed.

\paragraph*{State consistency}


 
When there is a state update from a data-plane node (e.g. an interface goes down or the node moves) the end-node sends an update to the controller and notifies other interested data-plane nodes of the change (e.g. peers with which it is communicating). That way, both the controller and the data-plane peers have the most updated information and know how to reach the end-node. However, northbound policy changes (e.g. a modified group policy changes the interface priorities for a set of nodes) may also result in updated state on the NIB. For that reason, data-plane nodes should periodically re-request state that they have cached from the controller to make sure that they have the most updated information. In addition, this periodic re-request can be complemented with the controller keeping track of the requesters for a particular piece of state. This is especially useful when a state update comes from a northbound application since the controller can directly notify the data-plane nodes affected by the change. 




\paragraph*{System scalability and resiliency}


Since all the controller nodes are interchangeable and offer the same functionality, the controller can be scaled-out smoothly and presents no single point of failure. The coordination subsystem can detect if the number of controlees increases or if a controller nodes goes down, and automatically instantiate a new controller node to redistribute the load. This process is transparent to the data-plane nodes. From the southbound point of view, data-plane nodes will detect an overloaded (or down) controller node by the lack of acknowledgment of their southbound requests and will eventually switch to a different one. This automatically balances the load across different controller nodes. The decentralized and symmetric schema for the controller also encourages data-plane nodes to seamlessly move across controller nodes to use the one with less load/latency. Furthermore, data-plane nodes can have mechanism to ensure they are able to forward traffic while they are retrieving the state or in the event that they lose connection to the controller (see section \ref{data-plane}).




\subsection{Southbound}
\label{shoutbound}



\begin{figure}[!t]
\centering
\includegraphics[width=0.45\textwidth]{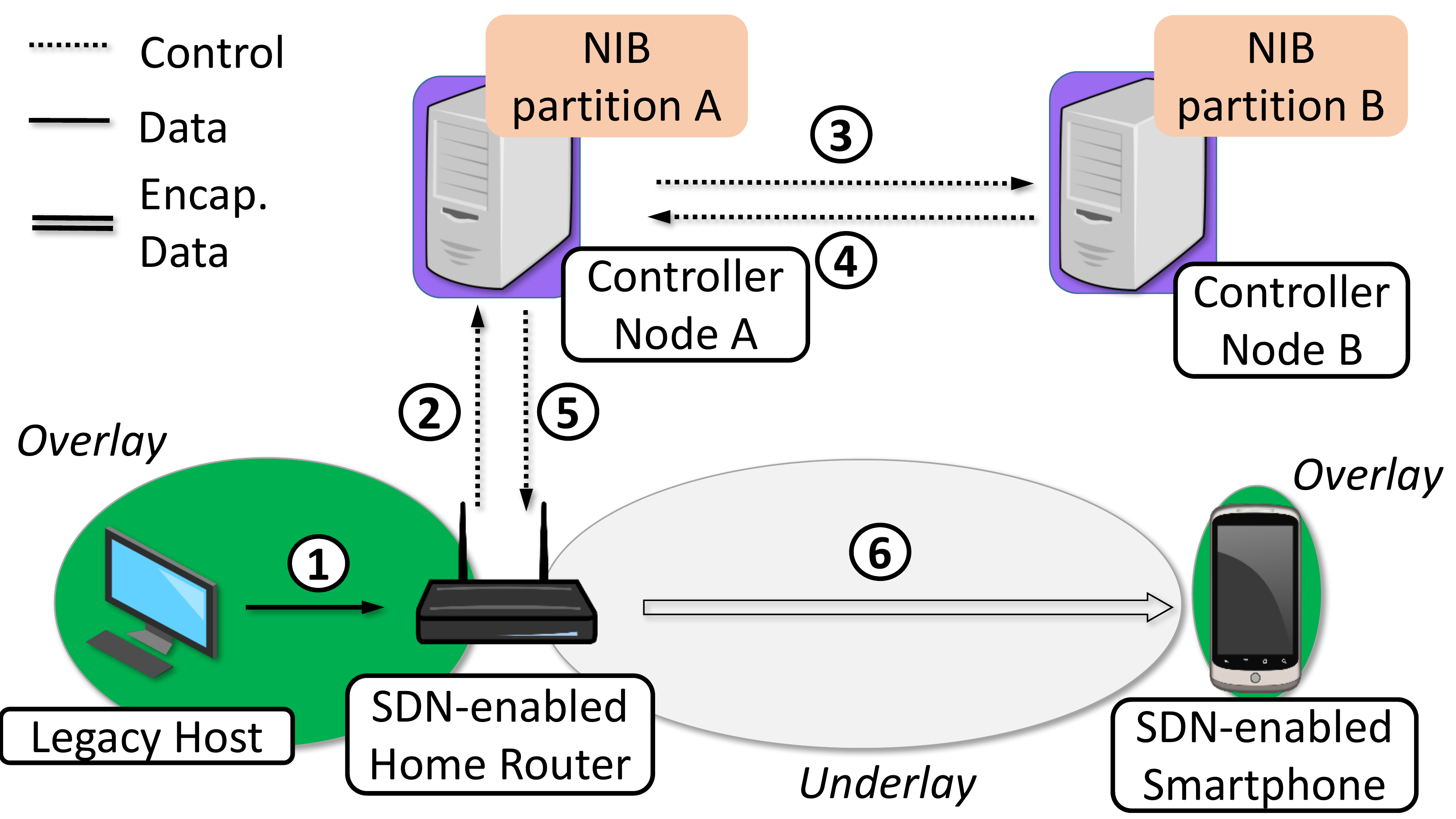}
\caption{\textbf{Southbound state retrieval} }
\label{fig:sb-state}
\end{figure}

The data-plane state is pulled on demand by end-nodes via a connectionless shouthbound protocol that uses a state interface to access the NIB. To illustrate southbound state retrieval and its sequence, Fig. \ref{fig:sb-state} shows an SDN-enabled home router encapsulating overlay traffic (1) towards an SDN-enabled smartphone. To do so, the home router requests the appropriate state to its nearest controller node via the southbound protocol (2). In the example depicted, the required state is not stored in NIB partition at the queried node and thus has to be obtained from another NIB partition (3)(4). The state is then sent to the home router (5) which uses it to encapsulate the overlay traffic through the underlay (6) towards the smartphone. 



All southbound queries will retrieve rendered state suitable for the requester node, although that may require one or more internal NIB queries (e.g. resolve a path between two end-nodes). To retrieve that state, generally, state requests include overlay traffic source and destination. Some use-cases may include also port or protocol information, or even details such as current geo-location or wireless channel(s) in use. Different use-cases may require extra information and NIB processing, but ideally the reply from the controller (that the data-plane node caches) should allow for simplified traffic classification. In many cases this may result in only source-destination based state (i.e. IP granularity).


\subsection{Data-plane}
\label{data-plane}





All data-plane nodes should be statically provisioned beforehand (e.g. via factory settings, configured by the user, etc) with the address of at least one controller node. This address can be a DNS name so it does not need to be fixed. Data-plane nodes use the controller address to retrieve bootstrap configuration via a remote configuration protocol (over a secure and reliable channel). Once the bootstrap is done, all future state retrieval will be done connectionless via southbound signaling. During the bootstrap, the controller provisions the nodes with their overlay address -or prefix-, an up-to-date list of controller nodes, and the rest of configuration they need for they data-plane operation (e.g. default egress policies, etc). This configuration should be periodically refreshed by the data-plane node. 

While the state is being retrieved, the data-plane node can choose to either buffer the traffic (if it has available resources) or rather encapsulate the traffic towards a re-encapsulation node. These nodes have larger hardware capabilities and can offer more buffering. Also, they can be provisioned per-group and therefore they can aggregate traffic, resulting in a high cache hit ratio. Since these nodes are static, the controller can always push updates to them to ensure they have the most recent state. These re-encapsulating nodes are deployed in different points of the network by the SDN provider and provisioned in the data-plane nodes via the bootstrap configuration. 








On the other hand, traffic encapsulation entitles to circumvent the traffic policies of the underlay and apply overlay policies instead. Different priorities and weights can be set for underlay locators and thus overlay routing can be enforced. In this sense, re-encapsulating nodes can be deployed in transit networks to help bypassing underlay out-of-scope policies. Re-encapsulating data-plane packets at these nodes enables to effectively detour traffic. To do so, re-encapsulating nodes can be added as intermediate hops in certain path policies.

Finally, if a data-plane node needs to communicate to a host outside the overlay namespace, it can redirect the traffic to a proxy node connected to the legacy Internet. The proxy node receives the traffic and takes care of decapsulating it and applying any NAT operation required (if the overlay traffic is not routable in the public Internet). Similarly, to the re-encapsulating nodes, these proxy nodes are configured in the data-plane nodes during bootstrap.

\section{Proof of concept}
\label{poc}



Following the design principles described above, we have built and evaluated the following proof-of-concept (PoC).

\subsection{Implementation}
\label{impl}

The implementation options considered for the PoC present a good trade-off between architectural adequateness and deployment feasibility. However, there are other options available and actual implementation choices will strongly depend on the particular use-case being considered. In particular for this PoC we propose the Group Based Policy (GBP)\footnote{https://wiki.openstack.org/GroupBasedPolicy} project as a starting point for the intent-driven norhtbound, since it is possible to leverage on its group-oriented syntax. For the controller we propose Opendaylight \cite{medved2014opendaylight}, a modular and extensible controller with support for multiple southbound and northbound protocols. For the NIB back-end we take advantage of Cassandra \cite{lakshman2010cassandra} since it is a general-purpose database that ensures constant latency times regardless of the number of entries stored \cite{rabl2012solving}. Besides, it offers an operations-per-second ratio linear to the number of database nodes \cite{rabl2012solving}. For the coordination module, we rely on ZooKeeper \cite{zookeeper}, while for the bootstrap protocol we use NETCONF (RFC6241). Both of them have wide support and are easy to integrate with other software pieces. To serve as southbound protocol we propose LISP (RFC6830), a protocol to map overlay identifiers to underlay network locators. We advocate for LISP in the southbound since it is connectionless, pull-based and overlay oriented \cite{rodriguez2015lisp}. Furthermore, there is already LISP support\footnote{https://wiki.opendaylight.org/view/OpenDaylight\_Lisp\_Flow\_Mapping:Main} in OpenDaylight. Finally, the architecture is agnostic to the specific format used for overlay encapsulation. For the PoC we choose VXLAN-GPE\footnote{https://tools.ietf.org/html/draft-ietf-nvo3-vxlan-gpe-05} since it can be easily integrated with LISP. To enable overlay capabilities at the end-nodes we use OpenOverlayRouter \cite{oor}, a multi-protocol and multi-platform (Android, Linux, OpenWrt) software router.

\subsection{Setup}

The PoC has been instantiated by means of allocating five virtual machines (Ubuntu 14.04, dual-core, 4GB of RAM) on the Amazon Web Services platform\footnote{http://aws.amazon.com}. The machines are geographically located at Europe (Ireland and Germany), US East Coast (Virginia) and US West Coast (California and Oregon). Table \ref{tab:pings} shows the average Round Trip Time (RTT) between all the PoC locations (measured over 1K ping iterations per pair). Each virtual machine hosts an instance of an OpenDaylight-like in-house controller (with NETCONF and LISP interfaces) that contains a node of a Cassandra cluster (ver. 3.1) storing the NIB. Additionally, to serve as end-node we deploy a laptop at our facilities at Barcelona running OpenOverlayRouter.

For the experiments, we consider the following use-case. End-nodes have different ingress and egress interface priorities depending to which group they belong. An end-node learns its egress priorities during bootstrap (via NETCONF) and has to query for other end-nodes ingress priorities on demand (via LISP). We artificially generate state and populate Cassandra for three different NIB sizes: 100K, 200K and 400K end-nodes. Respectively to the NIB size, we distribute the end-nodes across 100, 200 and 400 different groups. The per-node state is rendered in advance and already available for southbound retrieval.  For this PoC we use a single end-node since we focus on measuring latency associated with the state retrieval and update. For scenarios with a large number of nodes, other works have shown that the number of queries that Cassandra supports scales linearly with the number of database nodes and that the latency per-query is orthogonal to the number of queries \cite{rabl2012solving}.

\begin{table}[htbp!]
\begin{center}

\renewcommand{\arraystretch}{2.25}

\begin{tabular}{ r|c|c|c|c|c|c| } 

  \multicolumn{1}{c}{\rot[30]{}}
  & \multicolumn{1}{c}{\rot[30]{Barcelona}}
  & \multicolumn{1}{c}{\rot[30]{Ireland}}
  & \multicolumn{1}{c}{\rot[30]{Frankfurt}}
  & \multicolumn{1}{c}{\rot[30]{Virginia}}
  & \multicolumn{1}{c}{\rot[30]{California}}
  & \multicolumn{1}{c}{\rot[30]{Oregon}}
    \\ 
  \cline{2-7}
 Barcelona & - & 53 & 47 & 108 & 181 & 177   \\\cline{2-7}

 Ireland & 53 & - & 20 & 81 & 155 & 138  \\ \cline{2-7}

 Frankfurt & 47 & 20 & - & 89 & 166 & 165  \\ \cline{2-7}

 Virginia & 108 & 81 & 89 & - & 81 & 80   \\ \cline{2-7}

 California & 181 & 155 & 166 & 81 & - & 20   \\ \cline{2-7}

 Oregon & 177 & 138 & 165 & 80 & 20 & -   \\ \cline{2-7}
\end{tabular}
\end{center}

\caption{\textbf{Average RTT between PoC locations (in ms).} }
\label{tab:pings}
\end{table}



\subsection{Experiments}



\begin{figure}[!t]
\centering
\includegraphics[width=0.45\textwidth]{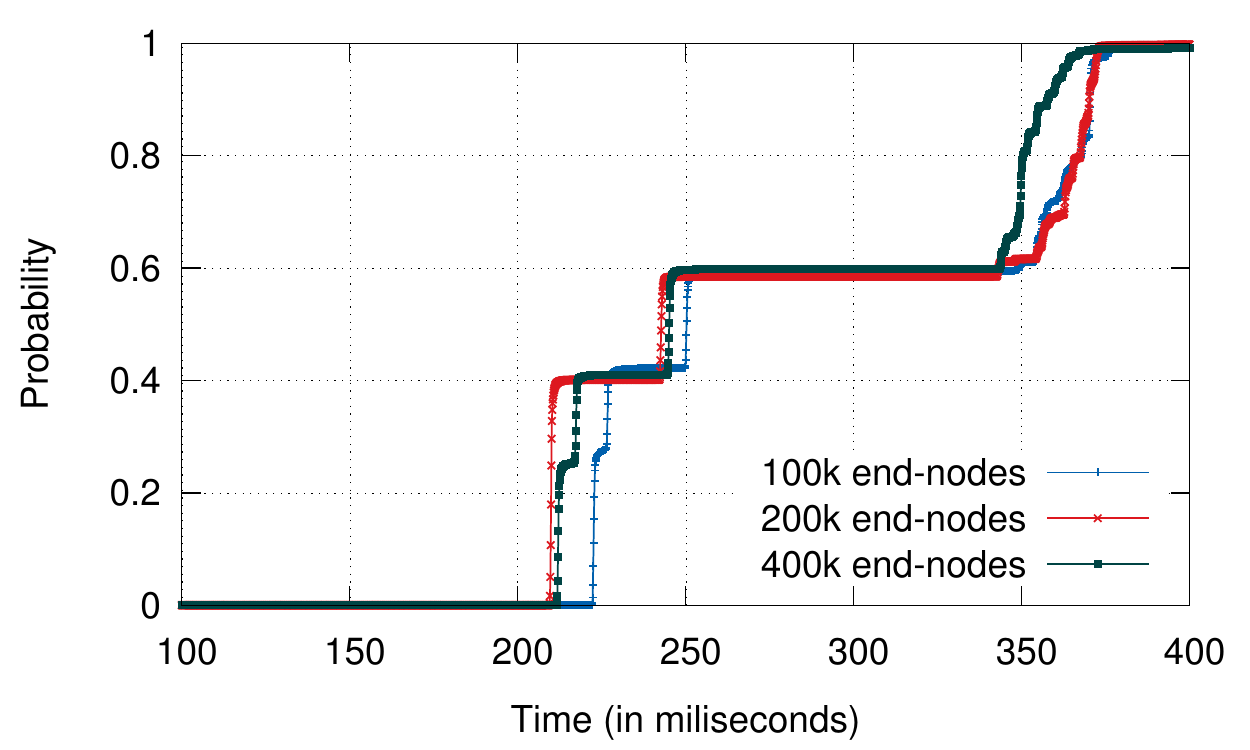}
\caption{\textbf{Southbound state retrieval} }
\label{fig:poc-a}
\end{figure}

\begin{figure}[!t]
\centering
\includegraphics[width=0.45\textwidth]{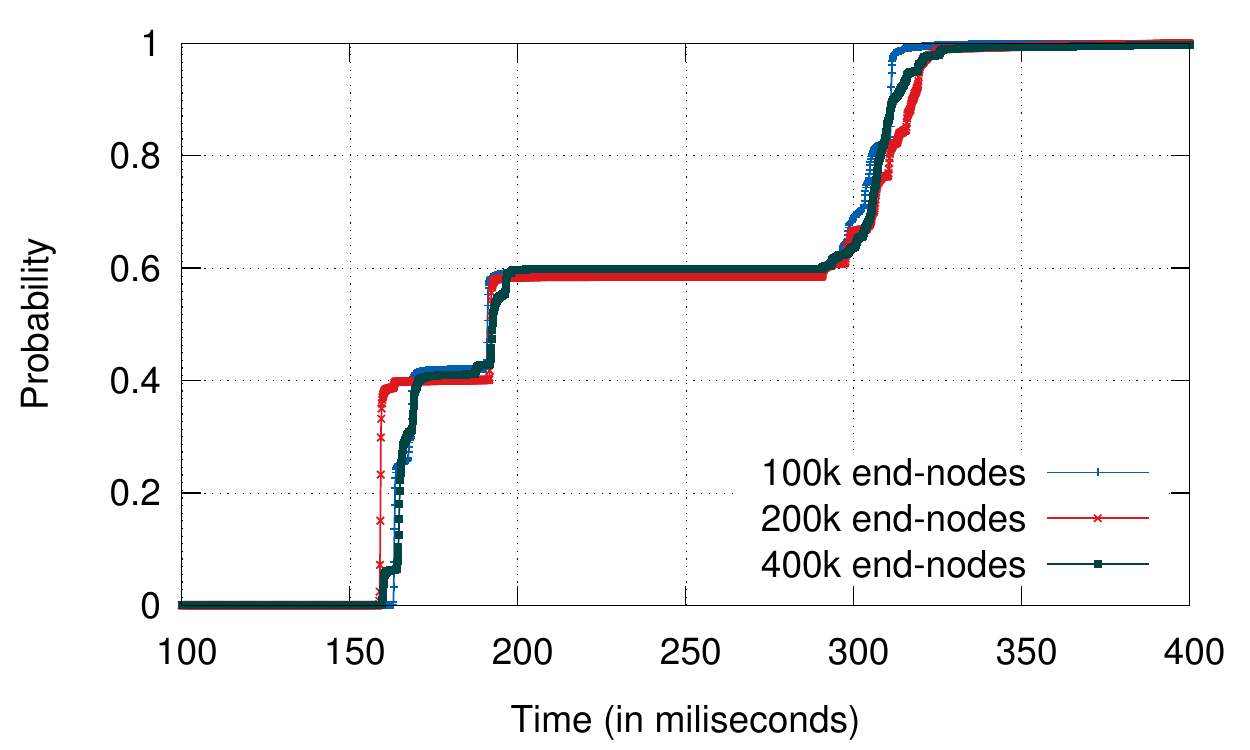}
\caption{\textbf{Northbound state update} }
\label{fig:poc-b}
\end{figure}

\begin{figure}[!t]
\centering
\includegraphics[width=0.45\textwidth]{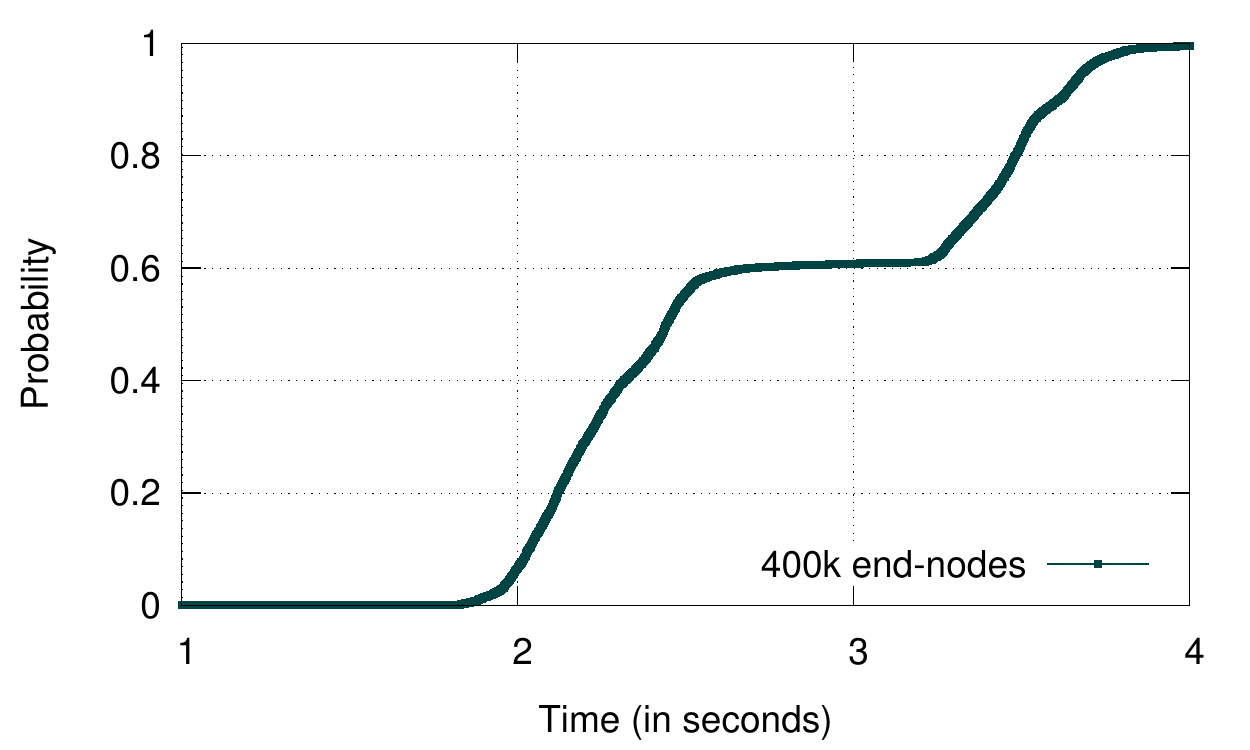}
\caption{\textbf{End-node bootstrap} }
\label{fig:poc-c}
\end{figure}

For a first experiment, we generate traffic in our end-node addressed to end-nodes stored in the NIB and measure the time the end-node takes to retrieve the appropriate routing policies from the NIB. We configure our end-node in Barcelona to use the controller node in Ireland. For each of the three different NIB sizes we generate 15K unique flows. We plot the latency results as Cumulative Distribution Functions (CDF) in Fig. \ref{fig:poc-a}, Fig. \ref{fig:poc-b} and Fig. \ref{fig:poc-c}. The results show that the latency is independent of the NIB size, which is coherent with the Cassandra results on \cite{rabl2012solving}. The figure also shows that the latency has a constant component (controller processing and controller to end-node transmission) and then a variable component, depending on from which Cassandra node the state has to be retrieved. Latency values tend to cluster around roughly three points, that can be associated with the different delays involved to retrieve state stored in either Europe, US West or US East. 



For a second experiment, we update the NIB state through the northbound and measure the time since the controller node (at Ireland) receives the northbound message until it sends the state-updated notification to the southbound. For each NIB size, we update state for 15K different end-nodes and plot the CDF in Fig. \ref{fig:poc-b}. The figure resembles Fig. \ref{fig:poc-a} since most of the latency is due again to the network transmission delay between the different Cassandra nodes. Northbound update time is measured directly at the controller and thus no latency is induced by the controller to end-node transmission. This makes Fig. \ref{fig:poc-b} to be shifted to the left around 50ms (Barcelona-Ireland RTT) when compared to Fig. \ref{fig:poc-a}. Besides, both Fig. \ref{fig:poc-a} and Fig. \ref{fig:poc-b} show minor latency variation across NIB sizes. This is due to the variation of network conditions over different iterations (since the PoC runs over the Internet) and to the randomness introduced during the process of generating traffic/updates.

Finally, we bootstrap the end-node and measure the time since it notifies its presence to the controller until the end-node acknowledges correct configuration installed. In this case, the end-node uses a random controller node on each iteration. We perform 7K iterations for the worst-case of 400K end-nodes and plot the CDF in Fig. \ref{fig:poc-c}. In this case, the latency depends not only on Cassandra, but also on the NETCONF exchange. 





\section{Related work}

There have been other works that discussed the interactions of end-nodes and SDN. Nevertheless, and to the best of our knowledge, this paper is the first ever to carefully explore all implications of a full SDN deployment of end-nodes. For instance, the authors of \cite{yap2012making} enable SDN within a mobile device to aggregate all its available interfaces. However, and contrary to our work, they do not consider that end-nodes can be transient, scattered, with low traffic locality and very numerous. Similarly, the authors of \cite{meneses2015extending} extend OpenFlow to bring SDN to end-nodes and complement existing approaches in the field of Software-defined Radio. They -intentionally- kept out the scope the complete architecture to support SDN-aware end-nodes. Similarly, meSDN \cite{lee2014mesdn} proposes a mobile extension for SDN to optimize wireless channel transmission on an existing SDN network. However, it does not consider devices connecting to legacy -non SDN- networks or transient devices that roam frequently. 

On the field of service-centric networking, Serval \cite{nordstrom2012serval} shows the value of enhancing the network capabilities of end-nodes. Despite its benefits, Serval is not designed to be an SDN solution and thus it does not offer centralized control or network programability. In a posterior work, OpenADN \cite{paul2012openadn} proposes a similar service-centric architecture but enhanced with an OpenFlow controller. However, the OpenADN framework focuses on switches at access networks and does not include devices at the very end of the network (e.g. smartphones). Consequently, it does not cover the challenges associated with controlees that are transient, heterogeneous and highly mobile.



\section{Conclusions}


This paper lays the foundations to further explore the implications of end-nodes in SDN deployments. The analysis concludes that the requirements of the end-nodes scenario differ from those of typical SDN deployments and thus need to be addressed via specific design guidelines. The design guidelines proposed in the paper are validated via a prototype that shows that an SDN architecture for end-nodes is feasible. 


As a final conclusion, it should be noted that in the scenario discussed, the control of the end-nodes is partially relinquished to the operator of the SDN infrastructure. This results in a tradeoff between the benefits that an SDN architecture could bring and the lack of control at the end-node.  In that sense, users have to decide if they are willing to trust their SDN providers in the same way that they trust, for instance, the manufacturer of their smartphone. Altough mechanisms like the ones described in \cite{privacy} can be used to protect the identity and location of users, further exploration of the privacy and security concerns of SDN for end-nodes remains as future work.

\end{document}